\documentclass[conference]{IEEEtran}
\usepackage[T1]{fontenc}
\usepackage{graphicx}
\graphicspath{{fig/}{./}}
\makeatletter\def\input@path{{fig/}{./}}\makeatother
\usepackage{inconsolata}
\usepackage{newtxmath}
\usepackage[nobreak]{cite}
\usepackage{url}
\usepackage{xspace}
\usepackage{eso-pic}

\usepackage{cleveref}
\crefname{figure}{Fig.}{Figs.}
\Crefname{figure}{Figure}{Figures}
\crefname{table}{Table}{Tables}
\crefname{section}{Section}{Sections}

\usepackage{framed}
\newcommand{\Conclusion}[1]{\begin{framed}\noindent #1\end{framed}}
\newcommand{\Heading}[1]{\textbf{#1.}}

\newcommand{\RQ}[1]{\textit{RQ}\textsubscript{#1}}
\newcommand{\defRQ}[2]{\expandafter\def\csname RQtext#1\endcsname{#2}}
\newcommand{\RQtext}[1]{\csname RQtext#1\endcsname}

\def\From{\mathit{from}}
\def\To{\mathit{to}}

\newcommand{\Ident}[1]{\texttt{#1}}
\newcommand{\Refactoring}[1]{\textsf{#1}}
\newcommand{\MSmath}{\mathsf{MS}}
\def\SS{\Refactoring{SS}\xspace}
\def\MSR{$\MSmath_\to$\xspace}
\def\MSL{$\MSmath_\gets$\xspace}
\def\MSRinv{$\MSmath_{\to\mathrm{Inv}}$\xspace}
\def\MSRret{$\MSmath_{\to\mathrm{Ret}}$\xspace}
\def\MSLinv{$\MSmath_{\gets\mathrm{Inv}}$\xspace}
\def\MSLret{$\MSmath_{\gets\mathrm{Ret}}$\xspace}

\def\SSfowler{\Refactoring{Slide Statements}\xspace}
\def\MSRfowler{\Refactoring{Move Statements into Function}\xspace}
\def\MSLfowler{\Refactoring{Move Statements to Callers}\xspace}
\def\SSL{\Refactoring{Slide Statements}\xspace}
\def\MSRinvL{\Refactoring{Move Statements into Method beyond Invocation}\xspace}
\def\MSRretL{\Refactoring{Move Statements into Method beyond Return}\xspace}
\def\MSLinvL{\Refactoring{Move Statements to Callers beyond Invocation}\xspace}
\def\MSLretL{\Refactoring{Move Statements to Callers beyond Return}\xspace}

\newcommand{\MEmath}{\mathsf{ME}}
\def\MER{$\MEmath_\to$\xspace}
\def\MERL{\Refactoring{Move Expression from Parameter}\xspace}
\def\MwH{\Refactoring{Move with Holes}\xspace}

\newcommand{\Meta}[1]{$\langle$\textit{#1}$\rangle$}
\newcommand{\etal}{\textit{et al.}\xspace}

\newcommand{\eg}{\textit{e.g.}\xspace}
\newcommand{\Code}[1]{\texttt{#1}}
\newcommand{\Project}[1]{\textit{#1}}

\begin{document}

\title{Formalizing and Automating\\Fine-Grained Move Refactorings Across Methods}
\author{%
  \IEEEauthorblockN{Kota Yasuhara}
  \IEEEauthorblockA{\textit{School of Computing}\\\textit{Institute of Science Tokyo}\\Tokyo, Japan\\yasuhara@se.comp.isct.ac.jp}
  \and
  \IEEEauthorblockN{Shinpei Hayashi}
  \IEEEauthorblockA{\textit{School of Computing}\\\textit{Institute of Science Tokyo}\\Tokyo, Japan\\hayashi@comp.isct.ac.jp}
}

\maketitle
\pagestyle{plain}
\thispagestyle{plain}

\AddToShipoutPictureBG*{%
  \AtPageLowerLeft{\raisebox{18.6mm}{%
    \hspace{\dimexpr\oddsidemargin+1in\relax}%
    \fbox{\parbox[b]{\dimexpr\textwidth-2\fboxsep-2\fboxrule\relax}{\tiny
      \copyright~2026 IEEE. Personal use of this material is permitted.
      Permission from IEEE must be obtained for all other uses, in any current or future media,
      including reprinting/republishing this material for advertising or promotional purposes,
      creating new collective works, for resale or redistribution to servers or lists,
      or reuse of any copyrighted component of this work in other works.}}}}}

\begin{abstract}
Developers use automated Move refactorings to improve the modular structure of source code and the assignment of responsibilities.
Class- and method-level Move refactorings are automated in modern IDEs, but statement- and expression-level moves that adjust method boundaries remain largely unautomated.
We formalize five variants of \Refactoring{Move Statement} refactoring as preconditions and steps grounded in four basic conditions covering data reachability, execution count, side effects, and syntactic constraints required for compilation, of which all but the side-effect condition are checked statically.
Combined with existing techniques, this also yields finer-grained moves of expressions and partial expressions.
We further refine the formalization iteratively against a real project, deriving twenty additional preconditions and steps that handle Java syntactic diversity in practice.
We evaluate applicability and compilability on ten projects, and behavior preservation in a case study on one of them: \Refactoring{Move Statement} refactorings yield compilable code in 93.3--97.0\% of applicable cases, and the case study shows that the observed behavioral changes stem from side-effect reordering left to developer judgment, not from defects in the statically checked conditions.
\end{abstract}

\section{Introduction}\label{s:introduction}

Refactoring has been widely used to improve the internal design of programs and continually improve the quality of modularization\cite{Fowler2018}.
Because maintenance dominates software development costs\cite{Pressman2009SoftwareEngineeringPractitioner, Glass2001FactsSoftwareEngineering} and program complexity tends to increase as development continues\cite{Lehman1979LawsProgramLifecycle}, improving the quality of modularization is essential to keep programs from becoming overly complex\cite{Stevens1974StructuredDesign}.
Refactoring remains a substantial part of everyday development: developers report spending about 10\% of their working time on refactoring\cite{Kim2014RefactoringChallengesAndBenefitsMicrosoft}, and recent large-scale studies confirm that it is practiced routinely\cite{Golubev2021OneThousandAndOneStoriesRefactoring, Zhao2025RefactoringRevisited}.

The \Refactoring{Move Method} refactoring\cite{Fowler2018} improves the quality of modularization by adjusting class boundaries and assigning methods to appropriate classes\cite{Al2018RefactoringQualitySurvey,Silva2016WhyWeRefactor}.
Refactoring tools in Integrated Development Environments (IDEs) can apply it automatically and safely\cite{Eclipse, IntelliJ, NetBeans}, and several approaches recommend \Refactoring{Move Method} opportunities\cite{AlDallal2017PredictingMoveMethodOpportunities, Cui2022MoveMethodRMove, Fokaefs2007MoveMethodJDeodorant, Kurbatova2020MoveMethodPathBased, Terra2018MoveMethodJMove}.

Finer-grained move refactorings, beyond the granularity of methods, can play a significant role in improving modular structure.
Whereas method-level move refactorings are limited to adjusting class boundaries, refactorings that move statement- or expression-level responsibilities between methods allow developers to flexibly redefine method boundaries and achieve more balanced responsibility assignments.
They are actively performed in real-world projects: the refactoring detection tool RefactoringMiner\cite{Tsantalis2018RefactoringMiner,Tsantalis2022RefactoringMiner2} identifies a refactoring named \Refactoring{Localize Parameter}, which converts a parameter into a local variable inside the callee method by moving the corresponding statement or expression from the call site into the callee.
A recent empirical study by Zhao \etal\cite{Zhao2025RefactoringRevisited} reports that \Refactoring{Localize Parameter} is performed on average 31.56 times per project, suggesting that developers frequently perform fine-grained move refactorings in practice.
Despite this demand, the foundations for formalizing and automating such refactorings remain insufficient.

Fowler defines several statement-level move refactorings\cite{Fowler2018}, but his informal catalog entries lack the precise preconditions and steps needed for automated, behavior-preserving application.
Sasaki \etal formalized the conditions for moving statements to different locations within the same method\cite{Sasaki2014SlideStatements}, but their approach cannot move statements to different methods.
Moreover, refactoring techniques for moving expressions have not been systematically established.
\Refactoring{Extract Local Variable} extracts an expression into a statement and \Refactoring{Extract Parameter} moves an expression or a variable declaration to the callers as a parameter, but no technique moves a computation into the callee by turning a parameter into a variable declaration.

\begin{figure*}[tb]\centering
  \nocite{Fowler2018,Sasaki2014SlideStatements,Tsantalis2022RefactoringMiner2,Chi2023ExtractLocalVariable,ExtractParameter}%
  \includegraphics{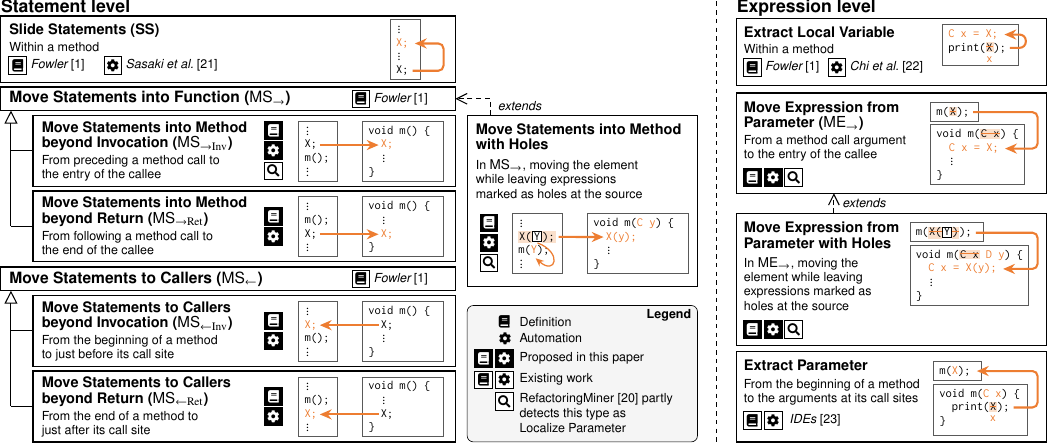}
  \caption{Fine-grained move refactorings addressed in this paper.}\label{f:overview}
\end{figure*}

In this paper, we propose a systematic and automated approach for fine-grained move refactorings, including \Refactoring{Move Statement} and \Refactoring{Move Expression}, as shown in \cref{f:overview}.
When moving statements across methods, four basic conditions must be satisfied to preserve program behavior: data references, execution counts, side effects and their ordering, and syntactic and structural constraints required for compilation.
To enforce these conditions, we statically analyze control flow and data dependencies, and define and automate the preconditions and steps of refactorings: all conditions but the side-effect one are guaranteed statically, whereas changes caused by side-effect reordering are deliberately left to developer judgment.

This paper focuses on the \emph{mechanics} of fine-grained move refactorings: deciding whether a given move satisfies the statically checked applicability conditions and performing the transformation automatically.
Deciding \emph{which} statements should move \emph{where}, a judgment that involves the conceptual cohesion of the code, is a complementary recommendation problem, just as an \Refactoring{Extract Method} engine leaves the choice of what to extract to its user.
Statically checked and automated mechanics are a prerequisite for any such recommender: a recommended move is useless if applying it breaks the program.

The recent rise of large language model (LLM)-based coding assistants reinforces rather than diminishes the need for formalized refactoring automation.
LLMs edit code probabilistically and offer no guarantee of behavior preservation, whereas a formalized engine checks explicit conditions statically by construction, runs locally at interactive speed, and explains a rejection in terms of the violated precondition.
The two are complementary rather than competing: recent \Refactoring{Move Method} assistants already combine LLM-based judgment about where code should live with static analysis that applies the change safely\cite{Bellur2025MMAssist, Zhang2025MoveMethodLLM}.
Our formalization supplies this statically checked execution layer at the statement and expression granularity, so that once an LLM or a developer decides that a statement belongs in a callee, the application can be delegated to a deterministic engine.
This division of labor becomes more valuable as agentic workflows chain many small edits, where per-step static checks keep errors from compounding.
Once automated with such guarantees, fine-grained moves can also be composed at scale: this work is thus a first step toward automating responsibility-assignment refactoring that optimizes the assignment of statements to methods across a system\cite{Mkaouer2015RemodularizationPackageNSGA-III, Amarjeet2017RemodularizationPackageHarmonySearch, Bright2019RemodularizationPackageColonyOptimization}.

The main contributions of this paper are as follows.
\begin{itemize}
    \item We formalize fine-grained move refactorings, including five \Refactoring{Move Statement} and three \Refactoring{Move Expression} variants, of which four and one are newly formalized, respectively, grounded in four basic conditions covering data reachability, execution count, side effects, and syntactic constraints required for compilation.
    \item We propose an iterative refinement methodology that adds preconditions or steps based on failures observed on a real project, yielding twenty additional rules.
    \item We empirically evaluate the formalization on ten projects, reporting applicability, compile success rates of 93.3--97.0\%, a test-based case study, and a cross-operation taxonomy of compile failures.
\end{itemize}

\section{Motivation}\label{s:motivatingExample}

\begin{figure}[tb]\centering
    \includegraphics{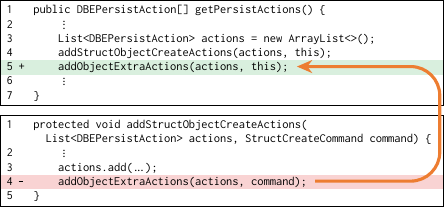}
    \caption{\Refactoring{Move Statement} refactoring in \textit{dbeaver}.}\label{2MoveStatementExaple}
\end{figure}

\Cref{2MoveStatementExaple} shows a \Refactoring{Move Statement} refactoring in \textit{dbeaver}\footnote{\url{https://github.com/dbeaver/dbeaver/commit/96a2ac2}}, where the call to \Ident{addObjectExtraActions} was moved from \Ident{addStructObjectCreateActions} to its caller \Ident{getPersistActions}.
If the callee method is invoked only from this caller, the move does not change the executed instructions and preserves the observable behavior.

Currently, developers must perform this refactoring manually, which risks inadvertently breaking program behavior\cite{Ge2012ManualRefactoringIsBad}: the argument \Ident{command} is defined inside the \Ident{addStructObjectCreateActions} method and cannot be referenced from \Ident{getPersistActions}, so developers must manually replace it with \Ident{this}, which references the same data.

\begin{figure}[tb]\centering
  \includegraphics{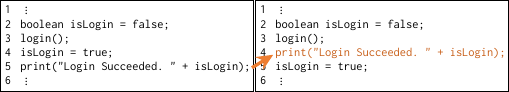}
  \caption{Example of a \Refactoring{Move Statement} that changes program behavior.}\label{3DataDependence}
\end{figure}

Automating \Refactoring{Move Statement} refactorings safely requires formalizing preconditions and steps that prevent the introduction of bugs.
In the login-processing code of \cref{3DataDependence}, if the statement \Ident{print("Login Succeeded. " + isLogin);} on Line 5 is moved between Lines 3 and 4, the printed value of \Ident{isLogin} changes from \Ident{true} to \Ident{false}, changing the program's observable behavior.

Fowler defines several \Refactoring{Move Statement} refactorings\cite{Fowler2018}: \SSfowler (\SS hereafter) moves statements within the same method.
Placing related operations, such as variable declarations and their uses, together improves code readability and prepares for other refactorings.
\MSRfowler (\MSR hereafter)\footnote{Although Fowler names this refactoring at the function level (\emph{into Function}), our operations are named for methods (\emph{into Method}) because of their focus on Java.} moves statements to a method called by their enclosing method, and \MSLfowler (\MSL hereafter), conversely, moves statements to the methods that call their enclosing method.
These refactorings adjust method boundaries, improving modularization at the statement level.

However, Fowler's definitions are not sufficient for automated, behavior-preserving refactoring: no preconditions are described, the steps mention only cutting and pasting code with no mechanism for resolving variable references across methods, and behavior equivalence is checked only by trial-and-error test runs.
Sasaki \etal formalized the conditions for reordering statements within the same method to improve readability\cite{Sasaki2014SlideStatements}, but their formalization targets only \SSfowler and cannot move statements across methods.

\begin{figure}[tb]\centering
  \includegraphics{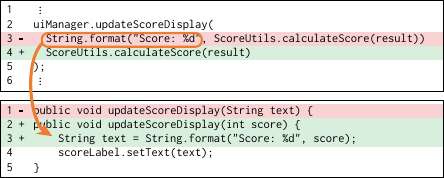}
  \caption{Example of \Refactoring{Move Expression} refactoring.}\label{2MoveExpressionExaple}
\end{figure}

This fine-grained movement approach can also be applied to partial computations contained in statements: in the motivating example of \cref{2MoveExpressionExaple}, which updates the score display on a game result screen, the call to the \Ident{updateScoreDisplay} method takes as an argument a string produced by calculating the score and then formatting it.
In terms of responsibility assignment, the score calculation should stay outside the \Ident{updateScoreDisplay} method while the string formatting should be inside it, so only the formatting part of the argument must be moved into the method.
However, such refactorings are not currently automated and are risky to perform manually.

\section{Defining Fine-Grained Move Refactorings}\label{s:approachMS}

This section formalizes the preconditions and steps of \Refactoring{Move Statement} refactorings based on Fowler's definitions\cite{Fowler2018}, refines the formalization iteratively against a real project, and derives additional refactorings by composing \Refactoring{Move Statement} with existing operations.

\Cref{f:overview} shows the fine-grained move refactorings we address.
\SSL (\SS) moves statements within the same method, similar to Fowler's definition, and is based on the formalization by Sasaki \etal\cite{Sasaki2014SlideStatements}.
\MSRinvL (\MSRinv hereafter) and \MSRretL (\MSRret hereafter) move statements to the beginning or end of the callee method, based on \MSRfowler; \MSLinvL (\MSLinv hereafter) and \MSLretL (\MSLret hereafter) move statements to immediately before or after the method call in the caller, based on {\MSLfowler}.
All but {\SS} are newly formalized in this paper.

We also apply Fowler's \Refactoring{Move Statement} concept to the expression level and address three \Refactoring{Move Expression} refactorings: \Refactoring{Extract Local Variable}\cite{Chi2023ExtractLocalVariable} corresponds to {\SSfowler}, {\MERL} ({\MER}) to {\MSRfowler}, and \Refactoring{Extract Parameter}\cite{ExtractParameter} to {\MSLfowler}.
The first and the last are based on existing operations, whereas {\MER} is new in this paper.

Finally, we newly define \Refactoring{Move with Holes}, in which the developer marks subexpressions of the moved element as holes: in {\MSR} and {\MER}, the marked holes stay in the source method while the rest moves to the callee.

In this section, we use {\MSRinvL} as an example to explain the preconditions and steps for \Refactoring{Move Statement} refactorings across methods.

\subsection{Considerations on Behavior Preservation}

In this paper, preserving behavior means that a method called from outside the program has the same return value and side effects before and after refactoring for any input.
We use Java as the target language for the formalization.

\begin{figure}[tb]\centering
  \includegraphics{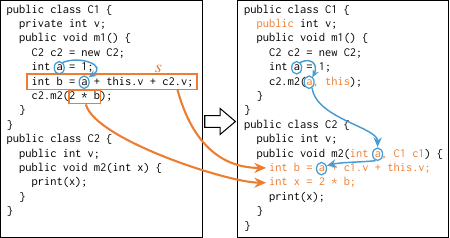}
  \caption{Application of \MSRinv.}\label{5MethodInvEx}
\end{figure}

{\MSRinv} takes as input a Java project $p$, a set of statements $S$ belonging to method $m_{\From}$, and the target method $m_{\To}$, and moves all statements in $S$ to the beginning of $m_{\To}$ in order.
\Cref{5MethodInvEx} shows an example: the statement \texttt{int b = a + this.v + c2.v;} in method \Ident{m1} of class \Ident{C1} is moved to the beginning of method \Ident{m2} of class \Ident{C2}, which is called by \Ident{m1}.

Suppose that method $m_{\From}$ calls method $m_{\To}$.
When the following four basic conditions (BC1--4) hold, statements can be moved across methods while preserving behavior.
\begin{itemize}
  \item BC1: All statements can reference the same data before and after the move.
  \item BC2: The execution counts of statements and expressions do not change.
  \item BC3: Each statement's side effects and the order among interacting ones do not change.
  \item BC4: Syntactic elements (statements, methods, etc.) meet the syntactic and structural constraints for compilation.
\end{itemize}

If BC1 and BC2 are both satisfied, each statement reads and writes the same values and is executed the same number of times, so it produces the same side effects the same number of times.
BC3 additionally requires that the order among interacting side effects be preserved, so that reordering does not change observable behavior.
BC4 prevents compilation errors from syntactic or structural violations.
Thus, we argue that satisfying BC1--4 preserves the program's behavior.

Our formalization is designed so that BC1, BC2, and BC4 are checked statically; \cref{s:RQ2} reports the residual cases in which the implementation falls short of this design.
BC3 is deliberately left to developer judgment, since a complete static check of side-effect ordering would be overly conservative in practice\cite{Liu2012InitialRefactoringTactics}, as discussed below.

This argument that BC1--4 jointly preserve behavior is an informal soundness sketch rather than a formal proof, following the convention of precondition-based refactoring formalizations such as the \Refactoring{Move Method} formalization by Tsantalis and Chatzigeorgiou\cite{Tsantalis2009JDeodorantApproach}.
It does not model concurrency, reflection, or the evaluation order of argument expressions, and changes in exception scopes, which fall under the control-flow equivalence required by BC2, are not treated exhaustively.
Instead of extending the static analysis to these rare constructs, we complement the formalization empirically: the iterative refinement in \cref{s:refinement} and the evaluation in \cref{s:evaluation} expose it to the syntactic diversity of real projects.

The rest of this subsection discusses how each basic condition is addressed, using {\MSRinv} as an example.

\begin{figure}[tb]\centering
  \includegraphics{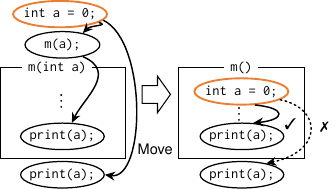}
  \caption{Data dependencies and data references.}\label{DDImage}
\end{figure}

\Heading{BC1: Same data references}
Consider the case where statement $s\in S$ defines variable $x$; data dependency analysis yields the statements that use this definition.
As shown in \cref{DDImage}, a statement inside the callee method that references $x$ via an argument can reference the same value after the move, whereas for a use of $x$ outside the callee method, the referenced definition may change or disappear.
We therefore require that a variable defined in $S$ be used only inside $S$ or in the arguments of the call to $m_{\To}$.

In \cref{5MethodInvEx}, the definition of variable \Ident{b} by statement $s$ is not used except in the call to method \Ident{m2}, so all statements can reference the same data before and after the move.
Note that this precondition is specific to {\MSRinv}: for {\MSLret}, which moves statements to immediately after the call in the caller, local variables used by $S$ must be passed back through the return value, so $S$ cannot be moved if it references multiple local variables.
In this way, per-operation preconditions ensure that all statements reference the same data.
Additionally, code transformations are needed to make the data references valid after the move; for example, a local variable used by a moved statement is added to the parameters of the callee method.

\Heading{BC2: Same execution counts}
To satisfy BC2, we adopt a conservative strategy that imposes two conditions: method $m_{\To}$ is called only once in method $m_{\From}$ and not called from anywhere else, and the set $S$ does not contain statements that jump outside $S$.
When both hold, the statements are executed exactly once per call of $m_{\To}$ in both positions, so the move does not change their execution count, as illustrated by the {\MSRinv} entry in \cref{f:overview}.
This strategy may reject safe moves, \eg, when every call site of $m_{\To}$ is preceded by identical statements, but it keeps the execution-count check simple and sound.
In addition, when adjustments for BC1 replace a variable reference with an expression, the expression's execution count may change; in such cases the formalization applies \Refactoring{Extract Local Variable} as needed so that BC2 continues to hold.

\Heading{BC3: Same side effects and ordering}
BC3 must be checked from two perspectives.
First, the side effect of a statement should not depend on its location: for example, moving a statement that prints a stack trace changes the stack-trace output.
Second, the order among statements whose side effects interact should not change: two \Code{add} calls appending elements to the same list produce different contents if their order is swapped.
In \cref{5MethodInvEx}, neither the moved statement nor the call to \Ident{m2} performs side-effecting operations, so BC3 is trivially satisfied.
In practice, however, side effects and their order do not always need to be preserved exactly: Liu \etal\cite{Liu2012InitialRefactoringTactics} report that developers do not always perform fully behavior-equivalent refactorings.
Building on this observation, we assume that minor changes such as stack-trace contents or debug-log order are often acceptable.
To avoid being overly conservative, we leave the judgment of such side-effect-related changes to developers; our preconditions do not enforce a complete check of side-effect ordering beyond what BC1 and BC2 cover.
A stricter treatment could integrate static purity and side-effect analyses\cite{Yang2014RevealingPurity, Yang2015PurityGuided} into the preconditions.
The run-time consequences of this design decision are examined in \cref{s:RQ3}.

\Heading{BC4: Syntactic and structural constraints}
BC4 prevents the moved code from violating syntactic or structural constraints required for compilation.
For example, if a statement throws a checked exception, that exception must be declared or handled in the enclosing method; assignments in the moved statements must respect \texttt{final} and definite-assignment rules at the destination; and the resulting class structure must remain syntactically valid.
Compilability is a central concern in refactoring formalization: Tsantalis and Chatzigeorgiou identify it as one of the main preconditions of \Refactoring{Move Method}\cite{Tsantalis2009JDeodorantApproach}, and 67.2\% of refactoring engine bugs manifest as compilation errors or engine crashes\cite{Wang2025RefactoringEngineBugs}.
We therefore introduce dedicated preconditions and steps for BC4, such as declaring or handling checked exceptions thrown by the moved statements in the destination method.
Note that steps that raise the visibility of accessed members or add required \texttt{import} declarations are syntactic adjustments in appearance but serve BC1: they restore the reachability of the data and members that the moved statements reference.
Raising the visibility of a member, however, weakens its encapsulation and thus stands in tension with the modularity improvement that motivates the move.
The current steps raise the visibility directly to \texttt{public}; more conservative alternatives, such as raising it only to the minimum required level or routing the access through accessors, are left as future work.

In \cref{5MethodInvEx}, the following five transformations are performed.
\begin{enumerate}
  \item Add the local variable \Ident{a} to the parameters.
  \item Pass \Ident{this} as an additional parameter and redirect member references through it.
  \item Replace the reference to the receiver \Ident{c2} of \Ident{m2} with \Ident{this}.
  \item Move \texttt{2 * b} in the arguments of \Ident{m2} into \Ident{m2}.
  \item Change the visibility of the private field \Ident{v} of class \Ident{C1} to public so that it can be referenced at the target.
\end{enumerate}

With the above preconditions and steps, as shown in \cref{5MethodInvEx}, statements can be moved across methods while preserving behavior, provided the developer confirms BC3.

\subsection{Formalization}

We show the catalog entry of {\MSRinv} as follows.
The catalog shows the final formalization after the iterative refinement described in \cref{s:refinement}.
The detailed formalization is available in the supplemental package\cite{supplementalPackage}.

\Heading{Name}
{\MSRinvL}

\Heading{Description}
\MSRinvL ({\MSRinv}) adjusts method boundaries toward more optimal responsibility assignment: it takes a Java project $p$, a set of statements $S$ in method $m_{\From}$, and the target method $m_{\To}$, and moves $S$ to the beginning of $m_{\To}$ in order.

\Heading{Precondition}
\begin{enumerate}
  \item The statements in $S$ are consecutive and are immediately followed by the call to $m_{\To}$ (by the definition of {\MSRinv}).
  \item Method $m_{\To}$ is called only once in method $m_{\From}$ and is not called from anywhere else (to satisfy BC2).
  \item Method $m_{\To}$ has a method body; this excludes abstract targets, whose implementation is determined dynamically.
  \item No statement $s \in S$ jumps outside $S$, \eg, via a \texttt{break} statement (to satisfy BC2).
  \item If statement $s \in S$ is a definition statement of variable $x$, variable $x$ is not used except inside $S$ or in the actual arguments of the call to $m_{\To}$ (to satisfy BC1).
\end{enumerate}

\Heading{Steps}
Let $A$ and $B$ be the receivers of $m_{\From}$ and $m_{\To}$.
\begin{enumerate}
  \item Move the statements in $S$ to the beginning of the body of method $m_{\To}$ by cut and paste.
  \item If statement $s \in S$ is a definition statement of variable $x$, remove the parameter of method $m_{\To}$ that corresponds to the use of variable $x$ in the arguments.
        If the removed actual argument is a direct reference to a variable, replace references to the removed formal parameter with references to variable $x$.
        Otherwise, add a variable declaration statement of the form \Meta{type} \Meta{formal parameter} = \Meta{actual argument}; immediately after statement $s$.
  \item In $S$, replace references to $B$ with \texttt{this}.
  \item If $S$ uses members of receiver $A$, add \texttt{this} to the actual arguments at the call to method $m_{\To}$ and add the corresponding declaration to the formal parameters of method $m_{\To}$.
        Then, redirect references to members of receiver $A$ in $S$ through the added parameter.
  \item Add local variables used in $S$ that are not members of receivers $A$ or $B$ to the formal parameters and actual arguments at the call to method $m_{\To}$.
  \item For methods, fields, and classes used by $S$, increase their visibility if they are not already accessible from the target class.
        Raising the visibility of a nested class also raises that of its enclosing classes.
  \item If $S$ references classes in other packages, add imports for them if not already imported.
\end{enumerate}

The other operations are formalized in the same format, and their complete catalog entries are available in the supplemental package\cite{supplementalPackage}.
The four cross-method operations differ mainly in where $S$ is inserted and how data crosses the method boundary.
\MSRret places $S$ at every exit of $m_{\To}$ and, when $S$ uses the call's result, requires the call form \Code{T x = m($\cdots$);} so that the use can be replaced with the return value.
\MSL moves $S$ to immediately before (\MSLinv) or after (\MSLret) every statement that calls $m_{\From}$, and replaces parameter references with the corresponding actual arguments.
It requires every call site to be an expression or variable declaration statement directly inside a block; \MSLret additionally restricts $m_{\From}$ to at most one tail \Code{return}, whose value is the only variable of $m_{\From}$ that $S$ may use.
Conditions uniform across the operations, as well as the constructor- and lambda-related restrictions, are given in the catalog.

\subsection{Iterative Refinement}\label{s:refinement}

The initial formalization, which follows the perspective of Sasaki \etal, mainly covers BC1 and BC2 but does not exhaustively cover the syntactic and structural constraints of BC4.
Since establishing a fully behavior-preserving formalization for every combination of Java constructs is impractical (even mature refactoring engines contain behavior-breaking bugs\cite{Wang2025RefactoringEngineBugs}), we instead refine it iteratively against a real project.

We choose \Project{mockito}, which is included in Defects4J\cite{Just2014Defects4J} and was used in the behavior-preservation evaluation of \Refactoring{Extract Local Variable}\cite{Chi2023ExtractLocalVariable}, as the target project.
For each of the four \Refactoring{Move Statement} operations, we enumerate all statement--method pairs that satisfy the current preconditions, apply each refactoring automatically, compile the project, run its test suite, and analyze every failure to update the formalization or its implementation.
The loop terminates when no failures remain or the residual failures are confirmed to lie outside the scope of the formalization, \eg, those caused by project-specific tooling.

For each failure, we use three guidelines to decide whether to forbid the offending element through a precondition or to absorb it through an additional step in the refactoring procedure.
First, an element whose compilation error cannot be avoided by relocation or by any existing step is excluded by a precondition; for example, an assignment to a \Code{final} field is excluded because the moved code would no longer compile at the destination.
Second, an element judged rare in practice is likewise excluded so as not to complicate the steps; for example, statements referencing variable-arity parameters (varargs) are simply rejected.
Third, an element that is common and amenable to code rewriting is absorbed by a step; for example, a checked exception is propagated by declaring it with \Code{throws} at the destination method instead of rejecting all statements that throw one.
Applying these guidelines to \Project{mockito} yielded twenty newly added preconditions and steps across the four operations.
\Cref{t:refinementRules} lists all twenty rules with the basic condition each serves; rows with multiple IDs bundle closely related rules, with hyphens connecting consecutive IDs.
In the table, P denotes a precondition and S a step.

\begin{table}[tb]\centering
  \caption{Rules Added through the Iterative Refinement}\label{t:refinementRules}
  {\scriptsize\setlength{\tabcolsep}{3.5pt}
  \newcommand{\thinhline}{\noalign{\hrule height 0.12pt}}
  \begin{tabular}{p{0.9cm}lp{6.6cm}}\hline
    ID & BC & Summary \\\hline
    P1.1.1 & BC1 & Statements containing \Code{super} calls are not moved to a different class \\\thinhline
    P1.2.2 & BC1 & Destination-receiver members referenced by moved statements are declared in the destination class or its superclasses \\\thinhline
    P1.3.5 & BC1 & Moved statements do not reference variable-arity parameters \\\thinhline
    P1.4.5 & BC1 & No moved uninitialized variable declaration is used outside the moved statements \\\thinhline
    P1.5.4, P1.5.6 & BC1 & Referenced types are resolvable in the destination class and are not type variables \\\thinhline
    P1.5.5 & BC1 & Moves from test code into production code carry no test-framework references \\\thinhline
    S1.5.3 & BC1 & Implicit references to \Code{static} methods are made explicit \\\thinhline
    P4.1.1 & BC4 & Signatures of overriding, overloaded, or variable-arity methods remain unchanged \\\thinhline
    S4.2.1 & BC4 & Checked exceptions unhandled at the destination are declared with \Code{throws} \\\thinhline
    P4.3.1 & BC4 & Moved statements do not assign to \Code{final} fields \\\thinhline
    P4.3.2 & BC4 & Constructor statements moved into a \Code{super()} call do not reference source-receiver members \\\thinhline
    P4.3.3--4 & BC4 & Neither the source method nor any of its callers is a constructor \\\thinhline
    P4.4.1--4 & BC4 & Every call to the source method is an expression or declaration statement directly inside a block in a method body, not a method reference \\\thinhline
    P4.4.5, S4.4.6 & BC4 & Expression-lambda bodies containing a call to the source method are convertible to, and rewritten into, block form \\\hline
  \end{tabular}}
\end{table}

\Cref{t:mockito} summarizes the compilation results on \Project{mockito} after refinement: the success rate ranges from 96.3\% to 98.8\% across the four operations, as expected on the project that drove the refinement; \cref{s:evaluation} reports the rates on the held-out projects.
The residual compilation failures fall into three categories: (a) false positives from a static analysis tool integrated into \Project{mockito}'s build, which rejects otherwise valid Java patterns (Infra); (b) flaky internal errors that disappear on re-execution (Infra); and (c) intrinsic limitations of the formalization, such as moving a statement immediately after a \Code{throw} statement (Limit).
Running \Project{mockito}'s test suite on the same instances served as the termination check of the refinement loop: 132 of 160 ({\MSRinv}), 131 of 165 ({\MSRret}), 431 of 571 ({\MSLinv}), and 207 of 409 ({\MSLret}) applications passed all tests.
The non-passing instances show the same pattern as the case study (\cref{s:RQ3}): they appear to be dominated by side-effect changes and reordering, which BC3 deliberately leaves to developer judgment, with the rest due to compile failures and documented limitations such as dynamic dispatch through abstract methods.

\begin{table}[tb]\centering
  \caption{Outcome of Refinement on \Project{mockito}}\label{t:mockito}
  {\footnotesize\begin{tabular}{l|rr} \hline
    Operation & Applied & Compiled (Rate) \\\hline
    \MSRinv & 160 & 158 (0.988) \\
    \MSRret & 165 & 161 (0.976) \\
    \MSLinv & 571 & 554 (0.970) \\
    \MSLret & 409 & 394 (0.963) \\\hline
  \end{tabular}}
\end{table}

\begin{table*}[tb]\centering
  \caption{Drag Sources and Drop Targets}\label{t:dnd}
  {\footnotesize\begin{tabular}{lll} \hline
    Refactoring & Drag source & Drop target \\\hline
    \SSfowler & Statements in method & The new location of the statement in the same method\\
    \MSRfowler (\MSRinv and \MSRret) & Statements in method & Call of target method\\
    \MSLfowler (\MSLinv and \MSLret) & Statements in method & Before or after the source method declaration\\
    {\Refactoring{Extract Local Variable}}\cite{Lee2013DragAndDropRefactoring} & Expression inside method & The new location of the statement in the same method\\
    \MERL & Actual argument & Call of target method\\
    {\Refactoring{Extract Parameter}}\cite{Lee2013DragAndDropRefactoring} & Expression inside method & Parameter list of the enclosing method signature\\\hline
  \end{tabular}}
\end{table*}
\begin{figure*}[tb]\centering
  \includegraphics[width=14cm]{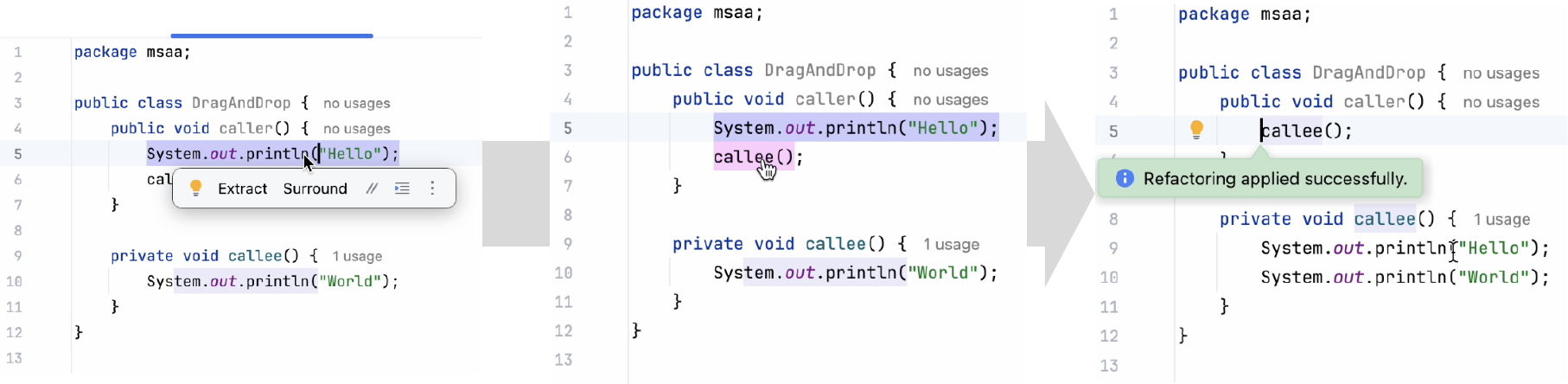}
  \caption{Applying \MSRinv\ by drag-and-drop: the developer selects a statement (left), drags it onto the call of the target method (middle), and the plugin checks the preconditions and applies the steps automatically (right).}\label{f:dndShot}
\end{figure*}

\subsection{Combination with Other Refactorings}\label{s:expansion}

By combining \Refactoring{Move Statement} refactorings with existing operations, we obtain finer-grained move refactorings.
We focus on two compositions: moving expressions across methods and excluding subexpressions from a move via Holes.

\Refactoring{Move Expression} refactorings relocate expressions within or across methods by composing \Refactoring{Move Statement} with expression-level operations.
\Refactoring{Extract Local Variable}\cite{Chi2023ExtractLocalVariable} achieves intra-method expression movement; it considers data dependencies and side effects.
\MERL\ moves an expression from a method-call argument to the beginning of the callee, by first extracting the expression with \Refactoring{Extract Local Variable} and then applying {\MSRinv}.
The opposite direction, from the callee to the call site, is provided by IntelliJ IDEA's \Refactoring{Extract Parameter}\cite{ExtractParameter}.
In our implementation, the preconditions of both constituent operations are checked in sequence before the composition is applied.

\MwH further generalizes {\MSR} and {\MER} by allowing the user to mark subexpressions as Holes that should remain in the source method.
The composition is: extract each Hole with \Refactoring{Extract Local Variable}, apply {\MSR} or {\MER}, and inline each extracted variable with \Refactoring{Inline Variable}.
This yields more flexible moves while reusing existing preconditions.

\section{Tool Support}\label{s:tool}

We implemented the formalized refactorings as a plugin for IntelliJ IDEA\cite{IntelliJ}, the most widely used Java IDE\cite{JrebelReport2025}.
The plugin checks preconditions and applies steps on the Program Structure Interface (PSI), which provides syntactic and data-flow analyses and rewrites code while preserving formatting.

Usability is a known barrier to refactoring tools: developers often refactor manually even when automated support exists\cite{Kim2014RefactoringChallengesAndBenefitsMicrosoft, Vakilian2012UseDisuseMisuse}.
To reduce the invocation cost, the plugin exposes the refactorings as Drag-and-Drop Refactoring\cite{Lee2013DragAndDropRefactoring}, in which a refactoring is invoked by dragging program elements onto a target.
\Cref{t:dnd} maps each refactoring to its drag source and drop target, and \cref{f:dndShot} shows {\MSRinv} applied by dragging a statement onto the call of the target method.
All static preconditions are checked automatically when the element is dropped, so the developer does not have to check them manually; the refactoring is either applied with the statically checked conditions satisfied or rejected with no change.
In addition, when a drag starts, the plugin highlights the drop targets of the cross-method operations ({\MSR} and {\MSL}) at which the static preconditions hold, so that the developer can see, before dropping, where the dragged statements can be moved.
The current implementation does not explicitly warn about potential side-effect reordering: BC3 is documented in the catalog as a developer-checked precondition, and surfacing it in the interface, \eg, as a drop-time warning or diff preview, is future work.
Evaluating the usability of this interface with developers is also future work.

\section{Evaluation}\label{s:evaluation}

\defRQ{1}{To what extent are \Refactoring{Move Statement} refactorings across methods applicable to statements?}
\defRQ{2}{Does the refactoring tool produce code without compile errors?}
\defRQ{3}{What behavioral changes remain after the transformation compiles successfully?}
\defRQ{4}{To what extent do the combined refactorings expand the applicable range?}

In this paper, we evaluate the applicability and behavior preservation of the newly formalized fine-grained move refactorings and answer the following research questions (RQs).
\begin{itemize}
  \item \RQ{1}: \RQtext{1}
  \item \RQ{2}: \RQtext{2}
  \item \RQ{3}: \RQtext{3}
  \item \RQ{4}: \RQtext{4}
\end{itemize}
The evaluation uses the plugin described in \cref{s:tool}.
We use ten projects from Defects4J\cite{Just2014Defects4J}, following prior refactoring evaluations\cite{Chi2023ExtractLocalVariable}: \Project{jfreechart}, \Project{commons-cli}, \Project{commons-codec}, \Project{commons-collections}, \Project{commons-compress}, \Project{jackson-core}, \Project{commons-jxpath}, \Project{commons-lang}, \Project{commons-math}, and \Project{joda-time}.
The ten projects together contain 67{,}509 methods, ranging from 1{,}043 (\Project{commons-cli}) to 11{,}700 (\Project{jfreechart}) per project.
Out of the 17 projects in Defects4J, we excluded \Project{mockito}, used to drive the iterative refinement (\cref{s:refinement}), and six projects that could not be built with the JetBrains Project System (JPS) used by our evaluation harness.
Detailed data are available in the supplemental package\cite{supplementalPackage}.

\subsection{\RQ{1}: Applicability}\label{s:RQ1}

We assess the applicability of the formalized refactorings from two angles: the number of methods that satisfy the per-method conditions, and the number of statement-level instances that satisfy all preconditions.

The per-method conditions are: for \MSR, the target is called exactly once and has a method body; for \MSL, every call site of the source method is a variable declaration or expression statement; for \MSLret, the source method additionally has at most one tail \texttt{return}.
Of the 67{,}509 methods in the dataset, 9{,}077 (13.4\%) satisfy the \MSR condition, 17{,}385 (25.8\%) the \MSLinv condition, and 14{,}557 (21.6\%) the \MSLret condition.
Although the percentages are modest, the absolute counts indicate a large supply of candidate methods.

For each operation we enumerate candidate (statement, target/source method) pairs, check whether \SS{} can shift the statement up to the method boundary, and check whether the cross-method preconditions hold.
\Cref{t:range} reports the totals.

\begin{table}[tb]\centering
  \caption{Applicability of \Refactoring{Move Statement} Refactorings}\label{t:range}
  {\footnotesize\begin{tabular}{l|rrr}\hline
    Operation & Targets & \SS-able (/Targets) & Applicable (/\SS-able) \\\hline
    \MSRinv & 37{,}160 & 9{,}948 (0.268) & 3{,}172 (0.319) \\
    \MSRret & 38{,}333 & 11{,}514 (0.300) & 2{,}654 (0.231) \\
    \MSLinv & 72{,}609 & 17{,}789 (0.245) & 13{,}825 (0.777) \\
    \MSLret & 58{,}239 & 17{,}943 (0.308) & 8{,}936 (0.498) \\\hline
  \end{tabular}}
\end{table}

The total number of applicable instances ranges from 2{,}654 (\MSRret) to 13{,}825 (\MSLinv).
These counts give an upper bound on the moves that satisfy the statically checked preconditions; whether a particular move improves the design is a separate question, as discussed in \cref{s:introduction}.
The \SS step is a strong filter: only 24.5--30.8\% of candidate statements can be reordered up to a method boundary, whereas 23.1--77.7\% of those that pass \SS also satisfy the cross-method preconditions.
Except for \MSRret, the \SS step keeps a smaller fraction of its candidates than the cross-method preconditions: intra-method reordering is the dominant constraint for most operations.

\Conclusion{%
  \Refactoring{Move Statement} refactorings offer thousands of applicable instances per operation, and for all but \MSRret, same-method reordering (\SS) is the dominant constraint.
}

\subsection{\RQ{2}: Compile Correctness}\label{s:RQ2}

Since 67.2\% of refactoring engine bugs manifest as compilation errors or engine crashes\cite{Wang2025RefactoringEngineBugs}, compile success is a meaningful first-line check, although it is only a necessary condition for behavior preservation.
For every applicable instance from \RQ{1} we apply the refactoring on a copy of the source tree and record whether the result compiles.

\Cref{t:compile} reports the totals.
Across all four operations the compile success rate is 93.3--97.0\%, with per-project rates ranging from 85.3\% to 100\%.

\begin{table}[tb]\centering
  \caption{Compilation Outcomes of \Refactoring{Move Statement}}\label{t:compile}
  {\footnotesize\begin{tabular}{l|rr}\hline
    Operation & Applied & Compiled (Rate) \\\hline
    \MSRinv & 3{,}172 & 3{,}019 (0.952) \\
    \MSRret & 2{,}654 & 2{,}477 (0.933) \\
    \MSLinv & 13{,}825 & 13{,}164 (0.952) \\
    \MSLret & 8{,}936 & 8{,}664 (0.970) \\\hline
  \end{tabular}}
\end{table}

We manually classified every compilation failure by the compiler diagnostic and mapped it to the violated basic condition.
We attribute a failure to the basic condition whose violation is the root cause; for example, a visibility or import error is attributed to BC1 when the moved statement can no longer reach the data or member it references.
We additionally assign each cause to one of four classes: \emph{implementation defects} (Defect), engine bugs to be fixed; \emph{documented limitations} (Limit), deliberately unsupported cases such as the import and argument-replacement boundaries; \emph{infrastructure failures} (Infra), flaky or project-specific build issues; and \emph{missing preconditions} (Missing PC), constructs for which the formalization lacks a rule, such as the scope-overlap precondition of \SS{} discussed below.
\Cref{t:failureTaxonomy} aggregates the top causes across all four operations, together with the violated basic condition and the assigned class.

\begin{table}[tb]\centering
  \caption{Causes of Compile Failures}\label{t:failureTaxonomy}
  {\scriptsize\setlength{\tabcolsep}{3.5pt}\begin{tabular}{l|l|l|r}\hline
    Cause (compiler diagnostic family) & BC & Class & Count \\\hline
    Variable shadowing on \SS{} inward & BC1 & Missing PC & 250 \\
    Import resolution limitation & BC1 & Limit & 206 \\
    Symbol not found (incl.\ rename collisions in \MSL) & BC1 & Defect & 202 \\
    Override visibility conflict & BC1 & Defect & 204 \\
    Argument-replacement limitation & BC1 & Limit & 82 \\
    Engine failure during application & N/A & Defect & 61 \\
    Final field/parameter assignment & BC4 & Missing PC & 54 \\
    Flaky internal engine errors & N/A & Infra & 43 \\
    Possibly uninitialized variable & BC4 & Missing PC & 23 \\
    Checked exception not caught/declared & BC4 & Missing PC & 2 \\
    Other rare diagnostics & various & various & 136 \\\hline
  \end{tabular}}
\end{table}

Three observations follow.
First, all individual causes with more than 100 occurrences are BC1 violations, which indicates that data reachability is the dominant residual concern.
Second, the override-visibility issue is concentrated in a single project (of its 204 occurrences, those in \MSR arise almost entirely in \Project{commons-compress}: 68 of 69 for \MSRinv and 114 of 114 for \MSRret), suggesting that some failures are project-idiosyncratic and that broader refinement on additional projects would help.
Third, BC4 violations (uninitialized variables, final-field assignment, checked-exception handling) are individually rare, but together they justify introducing BC4 as a separate condition: without its dedicated preconditions and steps, these cases would surface as compile errors.

Aggregated by class, implementation defects account for 467 of the 1{,}263 failures, missing preconditions for 329, documented limitations for 288, and infrastructure failures for 43.
The 61 engine crashes are tool defects rather than formalization gaps, and the remaining 136 failures scatter across infrequent diagnostic families.
Besides the documented limitations, engine crashes, and flaky errors, three high-impact problems remain, all violating BC1 (data references).
\begin{itemize}
  \item \textbf{Visibility-induced override breakage in \MSR{} (Defect).}
        When a moved statement references a method widened to \texttt{public}, the override chain in subclasses is not widened in lockstep, causing ``cannot override'' errors.
  \item \textbf{Variable double-definition in \MSLinv{} (Missing PC).}
        When two same-named variables live in disjoint scopes in the source method, moving a later declaration to the method head merges those scopes and produces a duplicate definition.
        The \SS{} formalization lacks the scope-overlap precondition forbidding this case; adding it and re-running the evaluation is immediate future work.
  \item \textbf{Rename-induced reference breakage in \MSL{} (Defect).}
        When a local variable and a method share an identifier, the name-clash avoidance logic renames not only the variable but also the call expression, breaking the reference; \eg, renaming \texttt{iterator} to \texttt{iterator1} also rewrites \texttt{iterator()} as \texttt{iterator1()}.
\end{itemize}

\Conclusion{%
  The refactoring tool produces compilable code for 93.3--97.0\% of applied instances, and the few high-impact failures all map to BC1 (data references), pointing to a small, focused remediation surface.
}

\subsection{\RQ{3}: Behavior Preservation: Case Study on \Project{commons-cli}}\label{s:RQ3}

To capture semantic regressions that survive the type-checker, we run the project's test suite after each successful application.
Because running the test suite for every applicable instance in all ten projects is prohibitively expensive, we conduct a case study on \Project{commons-cli}.
This $n{=}1$ design characterizes the residual failure modes that compilation cannot expose rather than estimating a population-level rate: the pass rates below are descriptive of this project and provide no evidence that this pattern occurs elsewhere.

\begin{table}[tb]\centering
  \caption{Test Outcomes on \Project{commons-cli}}\label{t:test}
  {\footnotesize\begin{tabular}{l|rr}\hline
    Operation & Applied & Tests passed (Rate) \\\hline
    \MSRinv & 127 & 34 (0.268) \\
    \MSRret & 57  & 31 (0.544) \\
    \MSLinv & 217 & 192 (0.885) \\
    \MSLret & 183 & 125 (0.683) \\\hline
  \end{tabular}}
\end{table}

\begin{table}[tb]\centering
  \caption{Causes of Test Failures on \Project{commons-cli}}\label{t:testFailureCauses}
  {\footnotesize\setlength{\tabcolsep}{3.5pt}\begin{tabular}{l|rrrr}\hline
    Cause & \MSRinv & \MSRret & \MSLinv & \MSLret \\\hline
    Reordering by \SS & 91 & 18 & 7 & 17 \\
    Reordering by the move itself & 0 & 4 & 3 & 28 \\
    Compile failure & 2 & 3 & 12 & 10 \\
    Positional side-effect change & 0 & 1 & 0 & 0 \\
    Dynamic dispatch limitation & 0 & 0 & 3 & 3 \\\hline
    Total failures & 93 & 26 & 25 & 58 \\\hline
  \end{tabular}}
\end{table}

\Cref{t:test} reports the test pass rate per operation on \Project{commons-cli}.
A manual triage of every failing instance attributed the failures to four cause classes; \cref{t:testFailureCauses} breaks down their distribution over all four operations:
  (i) reordering of side effects induced by \SS{}, the dominant cause for \MSRinv, \eg, individually movable \texttt{add} calls on the same collection whose observable order changes;
  (ii) side-effect changes induced by the cross-method move itself (mostly reordering, plus one positional change), which are more prominent for \MSL{} because the source method may be called multiple times;
  (iii) compile failures already counted in \RQ{2}; and
  (iv) implementation limitations around dynamic dispatch.
In terms of the failure classes of \cref{s:RQ2}, cause (iv) is a documented limitation and cause (iii) subsumes the classes analyzed there, whereas causes (i) and (ii) are consequences of the BC3 design decision rather than defects or missing preconditions.
Excluding causes (iii) and (iv), all observed behavioral changes stem from side effects and their ordering, which BC3 leaves to developer judgment by design (\cref{s:approachMS}); none was traced to a defect in the statically checked conditions (BC1, BC2, and BC4).
The low pass rate of \MSRinv (26.8\%) reflects this design decision, not a defect in the static checks.

\Conclusion{%
  On \Project{commons-cli}, aside from compile failures and tool limitations, every behavioral change stems from side-effect reordering that BC3 deliberately leaves to developer judgment; none traces to a defect in the statically checked conditions.
}

\subsection{\RQ{4}: Applicability of the Combined Refactorings}\label{s:RQ4}

We finally evaluate the applicability of \MER and \MwH, the refactorings obtained in \cref{s:expansion} by combining \MSR with \Refactoring{Extract Local Variable} and \Refactoring{Inline Variable}.
\MER is counted applicable when all preconditions hold and the actual argument is not a bare variable reference; \MwH when the moved element contains an extractable subexpression that is not a bare variable reference.
\Cref{t:rangeExpansion} reports the totals.

\begin{table}[tb]\centering
  \caption{Applicability of the Combined Refactorings}\label{t:rangeExpansion}
  {\footnotesize\begin{tabular}{l|rr}\hline
    & Applicable & w/ Holes (Rate) \\\hline
    \MSR & 5{,}826 & 4{,}232 (0.726) \\
    \MER & 1{,}107 & 195 (0.176) \\\hline
  \end{tabular}}
\end{table}

\MwH applies to 72.6\% of \MSR-applicable instances, indicating that statements moved across method boundaries frequently contain non-trivial subexpressions that could be left behind as holes.
\MER itself applies to 1{,}107 instances, fewer than for \MSR but still non-trivial; the smaller count reflects that \MER inherits the single-call-site requirement from \MSR while additionally restricting the argument shape.
Within \MER, \MwH applies to only 17.6\% of instances, indicating that arguments rarely contain a further extractable expression once they are themselves the move unit.

\Conclusion{%
  \MwH is broadly applicable on top of \MSR (72.6\%), while \MER applies at a smaller but practical scale (1{,}107 instances), with much rarer \MwH opportunities (17.6\%) inside it.
}

\subsection{Threats to Validity}

\subsubsection{Construct Validity}
We use compilation success and test pass rates as proxies for behavior preservation, but they do not capture every semantic deviation: subtle defects such as value changes via aliasing may pass both checks while still violating equivalence.
Soares \etal detect such deviations more systematically by generating tests that target the methods a refactoring changes\cite{Soares2013BehavioralTesting}, whereas we reuse each project's existing test suite, which exercises the refactored code less thoroughly.
For side-effect ordering, BC3 builds on the observation of Liu \etal\cite{Liu2012InitialRefactoringTactics} that developers do not always preserve behavior strictly; treating stack-trace differences as tolerable is our own assumption, and a stricter community standard would invalidate some applications we count as successful.
Our applicability counts measure where the refactorings \emph{can} be applied, not where they \emph{should} be applied; assessing the design-level desirability of individual moves, \eg, through cohesion metrics or developer studies, is outside our scope.

\subsubsection{Internal Validity}
The formalization of preconditions and application steps may contain flaws, and the evaluated tool may diverge from it due to implementation bugs, distorting the measured outcomes.
In \RQ{2}, we classified compilation failures and mapped them back to the basic conditions, but additional latent defects may not have surfaced as compile errors.
The failure-cause analysis in \RQ{3} relies on the authors' manual inspection and may include misclassifications.

\subsubsection{External Validity}
The evaluation targets ten open-source Java projects from Defects4J\cite{Just2014Defects4J}, so results may differ on closed-source projects or other domains.
The iterative refinement used a single project (\Project{mockito}); preconditions and steps may be biased toward its coding style, as the drop in compile success (96.3--98.8\% on \Project{mockito} vs.\ 93.3--97.0\% on the evaluation projects) and the \Project{commons-compress}-specific override-visibility failures suggest.
\RQ{3} is a case study on \Project{commons-cli}, selected because its test suite completes within our evaluation budget; it is a qualitative probe rather than a basis for quantitative generalization, and broader test-based evaluation is future work due to runtime constraints.
Smaller projects may have simpler code, so the case-study results may not generalize to larger projects.
Because the formalization targets Java, other languages may require different preconditions and steps due to differences in syntax and semantics.

\section{Related Work}\label{s:relatedwork}

Several remodularization approaches have been proposed to improve the quality of modularization by redistributing classes across packages or methods across classes.
Metaheuristic search techniques such as NSGA-III\cite{Mkaouer2015RemodularizationPackageNSGA-III}, harmony search\cite{Amarjeet2017RemodularizationPackageHarmonySearch}, and ant colony optimization\cite{Bright2019RemodularizationPackageColonyOptimization} have been applied to this problem, and their results must be realized as compositions of \Refactoring{Move Class} and \Refactoring{Move Method}, highlighting the importance of automated move refactorings.
However, whole-program remodularization often introduces thousands of scattered changes, which limits its practical adoption\cite{Hall2014RemodularisationIsHighCost}.
Our approach instead targets more localized, fine-grained moves that adjust method boundaries, providing statically checked, incremental code improvement opportunities.

As a less invasive alternative to whole-program remodularization, recommending individual \Refactoring{Move Method} operations has been studied to improve modularity at the method granularity.
Distance- and dependency-based approaches recommend moving a method to its most closely related class, using a distance metric between classes and methods\cite{Tsantalis2009JDeodorantApproach,Fokaefs2007MoveMethodJDeodorant}, attribute-, invocation-, and concept-based relationships\cite{AlDallal2017PredictingMoveMethodOpportunities}, or the similarity of dependency sets as in JMove\cite{Terra2018MoveMethodJMove}.
Machine-learning-based approaches capture the similarity between methods and classes from path-based code representations\cite{Kurbatova2020MoveMethodPathBased} or learned structural and semantic representations of code snippets as in RMove\cite{Cui2022MoveMethodRMove}.
More recently, MMAssist\cite{Bellur2025MMAssist} and MoveRec\cite{Zhang2025MoveMethodLLM} leverage LLMs combined with IDE features, static analysis, or deep learning for the recommendation.
While these approaches improve modularity at the method level, they cannot adjust method boundaries themselves; doing so requires moving statements or expressions across methods, as we do.

Statement-level move refactorings, the focus of this paper, were first cataloged by Fowler as \SSfowler, \MSRfowler, and \MSLfowler\cite{Fowler2018}, but only with informal descriptions lacking preconditions and steps precise enough for automated, behavior-preserving application.
Sasaki \etal formalized the reordering of statements within the same method based on data dependencies and control flow\cite{Sasaki2014SlideStatements}, imposing a Def-Use constraint preserving the order between a variable's definition and use, a Def-Def constraint preserving the order between definitions of the same variable, and an Escape constraint forbidding movement across statements that jump out of a block.
However, their formalization cannot move statements across method boundaries and does not consider the compilability of the resulting code.
In contrast, our approach formalizes four cross-method \Refactoring{Move Statement} variants and refines them iteratively against a real project.

\section{Conclusions and Future Work}\label{s:conclusion}
We formalized five \Refactoring{Move Statement} and three \Refactoring{Move Expression} variants of fine-grained move refactorings by grounding their preconditions and steps in four basic conditions over data reachability, execution count, side effects, and syntactic constraints for compilation.
We further refined the formalization iteratively against a real project, deriving twenty additional preconditions and steps that handle Java syntactic diversity in practice.
Our evaluation across ten Defects4J projects shows that the refactoring is applicable at a practical scale, that 93.3--97.0\% of applied cases compile successfully, and that the residual failures concentrate around BC1 (data reachability).
A test-based case study on \Project{commons-cli} further shows that the observed behavioral changes stem from side-effect reordering deliberately left to developer judgment by BC3, not from defects in the statically checked conditions.

Future work spans three directions.
The first widens the scope of the operations: supporting methods with multiple call sites, whose payoff can be bounded by counting how many rejected moves carry identical statements at every call site, and measuring how often the excluded constructs, such as concurrency and reflection, occur in practice.
The second strengthens the evidence: refining and testing on more projects than the single one used for each here, ablating the twenty refinement rules, and comparing against baselines such as compositions of existing IDE refactorings, LLM-based edits, and moves reproduced from commit histories.
The third turns these mechanics into developer-facing support: recommending which statements to move where, and evaluating the drag-and-drop interface with developers.

\Heading{Declaration of AI Usage}
The authors used Claude to improve presentation, including illustration, text readability, and language; they reviewed, edited, and take full responsibility for all content.

\Heading{Acknowledgments}
This work was partly supported by JSPS KAKENHI (JP26K02889, JP23K24823, JP26K02888, JP25K03102, JP25H01125, and JP24H00692).

\IEEEtriggeratref{40}
\bibliographystyle{IEEEtran}
\bibliography{references}

\begin{thebibliography}{10}
\providecommand{\url}[1]{#1}
\csname url@samestyle\endcsname
\providecommand{\newblock}{\relax}
\providecommand{\bibinfo}[2]{#2}
\providecommand{\BIBentrySTDinterwordspacing}{\spaceskip=0pt\relax}
\providecommand{\BIBentryALTinterwordstretchfactor}{4}
\providecommand{\BIBentryALTinterwordspacing}{\spaceskip=\fontdimen2\font plus
\BIBentryALTinterwordstretchfactor\fontdimen3\font minus
  \fontdimen4\font\relax}
\providecommand{\BIBforeignlanguage}[2]{{%
\expandafter\ifx\csname l@#1\endcsname\relax
\typeout{** WARNING: IEEEtran.bst: No hyphenation pattern has been}%
\typeout{** loaded for the language `#1'. Using the pattern for}%
\typeout{** the default language instead.}%
\else
\language=\csname l@#1\endcsname
\fi
#2}}
\providecommand{\BIBdecl}{\relax}
\BIBdecl

\bibitem{Fowler2018}
M.~Fowler, \emph{Refactoring: Improving the Design of Existing Code},
  2nd~ed.\hskip 1em plus 0.5em minus 0.4em\relax Addison-Wesley, 2018.

\bibitem{Pressman2009SoftwareEngineeringPractitioner}
R.~Pressman, \emph{Software Engineering: A Practitioner's Approach}.\hskip 1em
  plus 0.5em minus 0.4em\relax McGraw-Hill, Inc., 2009.

\bibitem{Glass2001FactsSoftwareEngineering}
\BIBentryALTinterwordspacing
R.~L. Glass, ``Frequently forgotten fundamental facts about software
  engineering,'' \emph{IEEE Software}, vol.~18, no.~3, pp. 110--112, 2001.
  [Online]. Available: \url{https://doi.org/10.1109/MS.2001.922739}
\BIBentrySTDinterwordspacing

\bibitem{Lehman1979LawsProgramLifecycle}
\BIBentryALTinterwordspacing
M.~M. Lehman, ``On understanding laws, evolution, and conservation in the
  large-program life cycle,'' \emph{Journal of Systems and Software}, vol.~1,
  pp. 213--221, 1980. [Online]. Available:
  \url{https://doi.org/10.1016/0164-1212(79)90022-0}
\BIBentrySTDinterwordspacing

\bibitem{Stevens1974StructuredDesign}
\BIBentryALTinterwordspacing
W.~P. Stevens, G.~J. Myers, and L.~L. Constantine, ``Structured design,''
  \emph{IBM Systems Journal}, vol.~13, no.~2, pp. 115--139, 1974. [Online].
  Available: \url{https://doi.org/10.1147/sj.132.0115}
\BIBentrySTDinterwordspacing

\bibitem{Kim2014RefactoringChallengesAndBenefitsMicrosoft}
\BIBentryALTinterwordspacing
M.~Kim, T.~Zimmermann, and N.~Nagappan, ``An empirical study of refactoring
  challenges and benefits at {Microsoft},'' \emph{IEEE Transactions on Software
  Engineering}, vol.~40, no.~7, pp. 633--649, 2014. [Online]. Available:
  \url{https://doi.org/10.1109/TSE.2014.2318734}
\BIBentrySTDinterwordspacing

\bibitem{Golubev2021OneThousandAndOneStoriesRefactoring}
\BIBentryALTinterwordspacing
Y.~Golubev, Z.~Kurbatova, E.~A. AlOmar, T.~Bryksin, and M.~W. Mkaouer, ``One
  thousand and one stories: {A} large-scale survey of software refactoring,''
  in \emph{Proceedings of the ACM Joint Meeting on European Software
  Engineering Conference and Symposium on the Foundations of Software
  Engineering (ESEC/FSE 2021)}, 2021, pp. 1303--1313. [Online]. Available:
  \url{https://doi.org/10.1145/3468264.3473924}
\BIBentrySTDinterwordspacing

\bibitem{Zhao2025RefactoringRevisited}
\BIBentryALTinterwordspacing
Y.~Zhao, X.~Li, J.~Guo, C.~Li, and L.~Nie, ``Refactoring revisited: An expanded
  study on refactoring practices,'' in \emph{Proceedings of the 25th
  International Conference on Software Quality, Reliability and Security (QRS
  2025)}, 2025, pp. 473--484. [Online]. Available:
  \url{https://doi.org/10.1109/QRS65678.2025.00054}
\BIBentrySTDinterwordspacing

\bibitem{Al2018RefactoringQualitySurvey}
\BIBentryALTinterwordspacing
J.~Al~Dallal and A.~Abdin, ``Empirical evaluation of the impact of
  object-oriented code refactoring on quality attributes: A systematic
  literature review,'' \emph{IEEE Transactions on Software Engineering},
  vol.~44, no.~1, pp. 44--69, 2018. [Online]. Available:
  \url{https://doi.org/10.1109/TSE.2017.2658573}
\BIBentrySTDinterwordspacing

\bibitem{Silva2016WhyWeRefactor}
\BIBentryALTinterwordspacing
D.~Silva, N.~Tsantalis, and M.~T. Valente, ``Why we refactor? {C}onfessions of
  {GitHub} contributors,'' in \emph{Proceedings of the 24th ACM SIGSOFT
  International Symposium on Foundations of Software Engineering (FSE 2016)},
  2016, pp. 858--870. [Online]. Available:
  \url{https://doi.org/10.1145/2950290.2950305}
\BIBentrySTDinterwordspacing

\bibitem{Eclipse}
``{Eclipse},'' \url{https://www.eclipse.org/}.

\bibitem{IntelliJ}
``{IntelliJ IDEA},'' \url{https://www.jetbrains.com/idea/}.

\bibitem{NetBeans}
``{NetBeans},'' \url{https://netbeans.apache.org/}.

\bibitem{AlDallal2017PredictingMoveMethodOpportunities}
\BIBentryALTinterwordspacing
J.~{Al Dallal}, ``Predicting move method refactoring opportunities in
  object-oriented code,'' \emph{Information and Software Technology}, vol.~92,
  pp. 105--120, 2017. [Online]. Available:
  \url{https://doi.org/10.1016/j.infsof.2017.07.013}
\BIBentrySTDinterwordspacing

\bibitem{Cui2022MoveMethodRMove}
\BIBentryALTinterwordspacing
D.~Cui, S.~Wang, Y.~Luo, X.~Li, J.~Dai, L.~Wang, and Q.~Li, ``{RMove}:
  Recommending move method refactoring opportunities using structural and
  semantic representations of code,'' in \emph{Proceedings of the 38th IEEE
  International Conference on Software Maintenance and Evolution (ICSME 2022)},
  2022, pp. 281--292. [Online]. Available:
  \url{https://doi.org/10.1109/ICSME55016.2022.00033}
\BIBentrySTDinterwordspacing

\bibitem{Fokaefs2007MoveMethodJDeodorant}
\BIBentryALTinterwordspacing
M.~Fokaefs, N.~Tsantalis, and A.~Chatzigeorgiou, ``{JDeodorant}: Identification
  and removal of feature envy bad smells,'' in \emph{Proceedings of the 23rd
  IEEE International Conference on Software Maintenance (ICSM 2007)}, 2007, pp.
  519--520. [Online]. Available:
  \url{https://doi.org/10.1109/ICSM.2007.4362679}
\BIBentrySTDinterwordspacing

\bibitem{Kurbatova2020MoveMethodPathBased}
\BIBentryALTinterwordspacing
Z.~Kurbatova, I.~Veselov, Y.~Golubev, and T.~Bryksin, ``Recommendation of move
  method refactoring using path-based representation of code,'' in
  \emph{Proceedings of the 42nd International Conference on Software
  Engineering (ICSE 2020) Workshops}, 2020, pp. 315--322. [Online]. Available:
  \url{https://doi.org/10.1145/3387940.3392191}
\BIBentrySTDinterwordspacing

\bibitem{Terra2018MoveMethodJMove}
\BIBentryALTinterwordspacing
R.~Terra, M.~T. Valente, S.~Miranda, and V.~Sales, ``{JMove}: A novel heuristic
  and tool to detect move method refactoring opportunities,'' \emph{Journal of
  Systems and Software}, vol. 138, pp. 19--36, 2018. [Online]. Available:
  \url{https://doi.org/10.1016/j.jss.2017.11.073}
\BIBentrySTDinterwordspacing

\bibitem{Tsantalis2018RefactoringMiner}
\BIBentryALTinterwordspacing
N.~Tsantalis, M.~Mansouri, L.~M. Eshkevari, D.~Mazinanian, and D.~Dig,
  ``Accurate and efficient refactoring detection in commit history,'' in
  \emph{Proceedings of the 40th International Conference on Software
  Engineering (ICSE 2018)}, 2018, pp. 483--494. [Online]. Available:
  \url{https://doi.org/10.1145/3180155.3180206}
\BIBentrySTDinterwordspacing

\bibitem{Tsantalis2022RefactoringMiner2}
\BIBentryALTinterwordspacing
N.~Tsantalis, A.~Ketkar, and D.~Dig, ``{RefactoringMiner} 2.0,'' \emph{IEEE
  Transactions on Software Engineering}, vol.~48, no.~3, pp. 930--950, 2022.
  [Online]. Available: \url{https://doi.org/10.1109/TSE.2020.3007722}
\BIBentrySTDinterwordspacing

\bibitem{Sasaki2014SlideStatements}
\BIBentryALTinterwordspacing
Y.~Sasaki, Y.~Higo, and S.~Kusumoto, ``Reordering program statements for
  improving readability,'' in \emph{Proceedings of the 17th European Conference
  on Software Maintenance and Reengineering (CSMR 2013)}, 2013, pp. 361--364.
  [Online]. Available: \url{https://doi.org/10.1109/CSMR.2013.50}
\BIBentrySTDinterwordspacing

\bibitem{Chi2023ExtractLocalVariable}
\BIBentryALTinterwordspacing
X.~Chi, H.~Liu, G.~Li, W.~Wang, Y.~Xia, Y.~Jiang, Y.~Zhang, and W.~Ji, ``An
  automated approach to extracting local variables,'' in \emph{Proceedings of
  the 31st ACM Joint European Software Engineering Conference and Symposium on
  the Foundations of Software Engineering (ESEC/FSE 2023)}, 2023, pp. 313--325.
  [Online]. Available: \url{https://doi.org/10.1145/3611643.3616261}
\BIBentrySTDinterwordspacing

\bibitem{ExtractParameter}
``{Extract parameter | IntelliJ IDEA Documentation},''
  \url{https://www.jetbrains.com/help/idea/extract-parameter.html}.

\bibitem{Bellur2025MMAssist}
\BIBentryALTinterwordspacing
A.~Bellur, F.~Batole, M.~R. Ullah, M.~Dilhara, Y.~Zharov, T.~Bryksin,
  K.~Ishikawa, H.~Chen, M.~Morimoto, T.~Hosomi, T.~N. Nguyen, H.~Rajan,
  N.~Tsantalis, and D.~Dig, ``Together we are better: {LLM}, {IDE} and semantic
  embedding to assist move method refactoring,'' in \emph{Proceedings of the
  41st IEEE International Conference on Software Maintenance and Evolution
  (ICSME 2025)}, 2025, pp. 1--13. [Online]. Available:
  \url{https://doi.org/10.1109/ICSME64153.2025.00046}
\BIBentrySTDinterwordspacing

\bibitem{Zhang2025MoveMethodLLM}
\BIBentryALTinterwordspacing
Y.~Zhang, Y.~Li, G.~Meredith, K.~Zheng, and X.~Li, ``Move method refactoring
  recommendation based on deep learning and {LLM}-generated information,''
  \emph{Information Sciences}, vol. 697, pp. 121\,753:1--17, 2025. [Online].
  Available: \url{https://doi.org/10.1016/j.ins.2024.121753}
\BIBentrySTDinterwordspacing

\bibitem{Mkaouer2015RemodularizationPackageNSGA-III}
\BIBentryALTinterwordspacing
W.~Mkaouer, M.~Kessentini, A.~Shaout, P.~Koligheu, S.~Bechikh, K.~Deb, and
  A.~Ouni, ``Many-objective software remodularization using {NSGA-III},''
  \emph{ACM Transactions on Software Engineering and Methodology}, vol.~24,
  no.~3, pp. 17:1--45, 2015. [Online]. Available:
  \url{https://doi.org/10.1145/2729974}
\BIBentrySTDinterwordspacing

\bibitem{Amarjeet2017RemodularizationPackageHarmonySearch}
\BIBentryALTinterwordspacing
Amarjeet and J.~K. Chhabra, ``Harmony search based remodularization for
  object-oriented software systems,'' \emph{Computer Languages, Systems \&
  Structures}, vol.~47, pp. 153--169, 2017. [Online]. Available:
  \url{https://doi.org/10.1016/j.cl.2016.09.003}
\BIBentrySTDinterwordspacing

\bibitem{Bright2019RemodularizationPackageColonyOptimization}
\BIBentryALTinterwordspacing
B.~G. {Varghese R}, K.~Raimond, and J.~Lovesum, ``A novel approach for
  automatic remodularization of software systems using extended ant colony
  optimization algorithm,'' \emph{Information and Software Technology}, vol.
  114, pp. 107--120, 2019. [Online]. Available:
  \url{https://doi.org/10.1016/j.infsof.2019.06.002}
\BIBentrySTDinterwordspacing

\bibitem{Ge2012ManualRefactoringIsBad}
\BIBentryALTinterwordspacing
X.~Ge, Q.~L. DuBose, and E.~Murphy-Hill, ``Reconciling manual and automatic
  refactoring,'' in \emph{Proceedings of the 34th International Conference on
  Software Engineering (ICSE 2012)}, 2012, pp. 211--221. [Online]. Available:
  \url{https://doi.org/10.1109/ICSE.2012.6227192}
\BIBentrySTDinterwordspacing

\bibitem{Liu2012InitialRefactoringTactics}
\BIBentryALTinterwordspacing
H.~Liu, Y.~Gao, and Z.~Niu, ``An initial study on refactoring tactics,'' in
  \emph{Proceedings of the 36th Annual IEEE Computer Software and Applications
  Conference (COMPSAC 2012)}, 2012, pp. 213--218. [Online]. Available:
  \url{https://doi.org/10.1109/COMPSAC.2012.31}
\BIBentrySTDinterwordspacing

\bibitem{Tsantalis2009JDeodorantApproach}
\BIBentryALTinterwordspacing
N.~Tsantalis and A.~Chatzigeorgiou, ``Identification of move method refactoring
  opportunities,'' \emph{IEEE Transactions on Software Engineering}, vol.~35,
  no.~3, pp. 347--367, 2009. [Online]. Available:
  \url{https://doi.org/10.1109/TSE.2009.1}
\BIBentrySTDinterwordspacing

\bibitem{Yang2014RevealingPurity}
\BIBentryALTinterwordspacing
J.~Yang, K.~Hotta, Y.~Higo, and S.~Kusumoto, ``Revealing purity and side
  effects on functions for reusing {Java} libraries,'' in \emph{Proceedings of
  the 14th International Conference on Software Reuse (ICSR 2014)}, 2015, pp.
  314--329. [Online]. Available:
  \url{https://doi.org/10.1007/978-3-319-14130-5_22}
\BIBentrySTDinterwordspacing

\bibitem{Yang2015PurityGuided}
\BIBentryALTinterwordspacing
------, ``Towards purity-guided refactoring in {Java},'' in \emph{Proceedings
  of the 31st IEEE International Conference on Software Maintenance and
  Evolution (ICSME 2015)}, 2015, pp. 521--525. [Online]. Available:
  \url{https://doi.org/10.1109/ICSM.2015.7332506}
\BIBentrySTDinterwordspacing

\bibitem{Wang2025RefactoringEngineBugs}
\BIBentryALTinterwordspacing
H.~Wang, Z.~Xu, H.~Zhang, N.~Tsantalis, and S.~H. Tan, ``Towards understanding
  refactoring engine bugs,'' \emph{ACM Transactions on Software Engineering and
  Methodology}, vol.~35, no.~5, pp. 138:1--55, 2026. [Online]. Available:
  \url{https://doi.org/10.1145/3747289}
\BIBentrySTDinterwordspacing

\bibitem{supplementalPackage}
\BIBentryALTinterwordspacing
K.~Yasuhara and S.~Hayashi, ``Artifact for ``{F}ormalizing and automating
  fine-grained move refactorings across methods'','' figshare, 2026. [Online].
  Available: \url{https://doi.org/10.6084/m9.figshare.32657553}
\BIBentrySTDinterwordspacing

\bibitem{Just2014Defects4J}
\BIBentryALTinterwordspacing
R.~Just, D.~Jalali, and M.~D. Ernst, ``{Defects4J}: {A} database of existing
  faults to enable controlled testing studies for {Java} programs,'' in
  \emph{Proceedings of the 23rd International Symposium on Software Testing and
  Analysis (ISSTA 2014)}, 2014, pp. 437--440. [Online]. Available:
  \url{https://doi.org/10.1145/2610384.2628055}
\BIBentrySTDinterwordspacing

\bibitem{Lee2013DragAndDropRefactoring}
\BIBentryALTinterwordspacing
Y.~Y. Lee, N.~Chen, and R.~E. Johnson, ``Drag-and-drop refactoring: Intuitive
  and efficient program transformation,'' in \emph{Proceedings of the 35th
  International Conference on Software Engineering (ICSE 2013)}, 2013, pp.
  23--32. [Online]. Available: \url{https://doi.org/10.1109/ICSE.2013.6606548}
\BIBentrySTDinterwordspacing

\bibitem{JrebelReport2025}
{JRebel}, ``2025 {Java} developer productivity report,''
  \url{https://www.jrebel.com/resources/java-developer-productivity-report-2025},
  2025.

\bibitem{Vakilian2012UseDisuseMisuse}
\BIBentryALTinterwordspacing
M.~Vakilian, N.~Chen, S.~Negara, B.~A. Rajkumar, B.~P. Bailey, and R.~E.
  Johnson, ``Use, disuse, and misuse of automated refactorings,'' in
  \emph{Proceedings of the 34th International Conference on Software
  Engineering (ICSE 2012)}, 2012, pp. 233--243. [Online]. Available:
  \url{https://doi.org/10.1109/ICSE.2012.6227190}
\BIBentrySTDinterwordspacing

\bibitem{Soares2013BehavioralTesting}
\BIBentryALTinterwordspacing
G.~Soares, R.~Gheyi, and T.~Massoni, ``Automated behavioral testing of
  refactoring engines,'' \emph{IEEE Transactions on Software Engineering},
  vol.~39, no.~2, pp. 147--162, 2013. [Online]. Available:
  \url{https://doi.org/10.1109/TSE.2012.19}
\BIBentrySTDinterwordspacing

\bibitem{Hall2014RemodularisationIsHighCost}
\BIBentryALTinterwordspacing
M.~Hall, M.~A. Khojaye, N.~Walkinshaw, and P.~McMinn, ``Establishing the source
  code disruption caused by automated remodularisation tools,'' in
  \emph{Proceedings of the 30th IEEE International Conference on Software
  Maintenance and Evolution (ICSME 2014)}, 2014, pp. 466--470. [Online].
  Available: \url{https://doi.org/10.1109/ICSME.2014.75}
\BIBentrySTDinterwordspacing

\end{thebibliography}

\end{document}